\documentclass[preprint,proceedings]{rmaa}

\usepackage{paralist}

\usepackage{psfrag,color}

\SetYear{2005}
\SetConfTitle{Triggering Relativistic Jets}

\title{The MHD collapsar model for GRBs: an inflow produces an outflow} 

\author{
  D. Proga \altaffilmark{1} }

\altaffiltext{1}{Department of Physics, University of Nevada, Las Vegas, USA}

\shortauthor{Proga}
\shorttitle{MHD collapsar model}

\fulladdresses{
\item Daniel Proga; Department of Physics, University of Nevada, 4505 South Maryland Parkway, Las Vegas, NV 89154, USA 
(\email{dproga@physics.unlv.edu}).
}

\listofauthors{D. Proga}
\indexauthor{D. Proga.}

\abstract{We present our recent results from numerical simulations
of a magnetized flow in the vicinity of a black hole in the context of 
the collapsar model for GRBs. The simulations show that after an 
initial transient, the flow settles into a complex convolution 
of several distinct, time-dependent flow components 
including an accretion torus, its corona and outflow,
an inflow and an outflow in the polar funnel.
We focus on studying the nature and connection between these components,
in particular between the inflows and related outflows. 
We find that rotational and MHD effects launch, accelerate, 
and sustain the outflows. We also find that an outflow can be formed even 
when the collapsing envelope has initially a very weak magnetic field and 
a very small angular momentum. Our main conclusion is that even 
for a relatively weak initial magnetic field and a slow rotation, 
a gravitational collapse of a stellar envelope can lead to formation  of
a very strong and very fast jet.}

\addkeyword{Accretion, Accretion disks}
\addkeyword{Gamma rays: bursts}
\addkeyword{Methods: numerical}
\addkeyword{MHD}
\addkeyword{Stars: winds, outflows}

\begin{document}

\def\LSUN{\rm L_{\odot}}
\def\MSUN{\rm M_{\odot}}
\def\RSUN{\rm R_{\odot}} 
\def\MSUNYR{\rm M_{\odot}\,yr^{-1}}
\def\MSUNS{\rm M_{\odot}\,s^{-1}}
\def\MDOT{\dot{M}}

\newbox\grsign \setbox\grsign=\hbox{$>$} \newdimen\grdimen \grdimen=\ht\grsign
\newbox\simlessbox \newbox\simgreatbox
\setbox\simgreatbox=\hbox{\raise.5ex\hbox{$>$}\llap
     {\lower.5ex\hbox{$\sim$}}}\ht1=\grdimen\dp1=0pt
\setbox\simlessbox=\hbox{\raise.5ex\hbox{$<$}\llap
     {\lower.5ex\hbox{$\sim$}}}\ht2=\grdimen\dp2=0pt
\def\simgreat{\mathrel{\copy\simgreatbox}}
\def\simless{\mathrel{\copy\simlessbox}}
\maketitle

\section{Introduction}
\label{sec:intro}

Gamma-ray burts (GRBs) are associated with the huge release of energy 
in a matter of seconds. The collapsar model is one of most promising 
scenarios to explain these as well as other properties of GRBs
(Woosley 1993; Paczy\'{n}ski 1998; MacFadyen \& Woosley 1999; 
Popham, Woosley \& Fryer 1999;MacFadyen, Woosley \& Heger 2001; 
Proga et al. 2003). In this scenario, the collapsed iron
core of a massive star accretes gas at a high rate ($\sim 1 \MSUNS$)
producing a large neutrino flux, a powerful outflow, and a GRB.
Many breakthroughs were made in studying GRBs. For example,
the association of long duration GRBs 
with stellar collapse was firmly confirmed
(Hjorth et al. 2003, Stanek et al. 2003). Nevertheless,
basic properties of the GRB central engine are quite uncertain.
This is because the physical conditions in the central engine 
are extreme and complex (e.g.,  gravitational field is very strong, 
the temperature and the  mass and energy densities are very high; 
large scale supersonic/relativistic flows and small scale turbulent flows
are physically connected).
Additionally, magnetic fields are most likely very important
in determining the properties of the central engine.

Effects of magnetic fields in the context of the collapsar model
have been studied by a few groups
(e.g., Mizuno, Yamada, Koide, \& Shibata 2004; Mizuno, et al. 2004;
De Villiers, Staff, \& Ouyed 2005). The main focus
of these studies is on 2 and 3 dimensional 
general relativistic magnetohydrodynamic (GR MHD) simulations of jets 
launched self-consistently from accretion disks orbiting Schwarzschild or
Kerr black holes. These are very important studies as they capture
a few of the key elements of the central engine and soon may
include more elements such as sophisticated equation of state and
neutrino physics.

Here, we present and discuss results from 2.5-dimensional,
magnetohydrodynamic (MHD) simulations of the collapsar model 
using pseudo-Newtonian potential. 
These simulations (see also Proga et al. 2003) are an extension of 
the work of Proga \& Begelman (2003, hereafter PB03) who studied 
MHD accretion flows onto a black hole (BH). 
In particular, the collapsar simulations include a realistic equation of 
state, photodisintegration of bound nuclei and cooling due to neutrino 
emission. The simulations presented here 
are also an extension of collapsar simulations by 
MacFadyen \& Woosley (1999), as they include very similar neutrino physics and 
initial conditions but are in the MHD instead hydrodynamical (HD) limit.

\section{Models} 
\label{sec:Models} 

Proga et al.'s simulation begins after the $1.7~\MSUN$ iron
core of a 25~$\MSUN$ presupernova star has collapsed and follows the
ensuing accretion of the $7~\MSUN$ helium envelope onto the central
black hole formed by the collapsed iron core.  
A spherically symmetric progenitor model is assumed, 
but the symmetry is
broken by the introduction of a small, latitude-dependent angular
momentum and a weak split-monopole  magnetic field. 
For more details, see Proga et al. (2003). We refer to the model presented
in Proga et al. as run A. We will also present
results from model~B which is a rerun of model A with
the numerical resolution increased by a factor of 2 in the 
latitudinal direction (see section 3.2). Run B has also a higher resolution
in the radial direction compared to run A; the resolution increase
is a function of radius and is of a factor of 19 at the smallest radii
and of a factor of 1.015 at the largest radii. 
Run B differs also from run A in the way we set up the initial conditions.
However, despite these differences the gross properties of the
solution during the later phase of the evolution are very similar
for both runs.
The details of run B are in Proga (2005; in preparation).

\section{Results}
\label{sec:results}

\subsection{Fiducial run}

Figure~1 shows time sequence of logarithmic density (top) and 
toroidal magnetic field maps (bottom) overplotted with the direction
of the poloidal velocity from run A during the early phase of the evolution.
The figure illustrates how after 
a transient episode of infall, lasting about 0.13 s, the gas with 
$l\approx 2 R_S c$ starts to pile up outside the black hole and to 
form a thick torus bounded by a centrifugal barrier near the rotation axis.  
Soon after the torus forms (i.e., within a couple of orbital times at
the inner edge), the magnetic field is amplified by shear and 
the magnetorotational
instability (MRI, e.g., Balbus \& Halwey 1991).
We note that the third possible mechanism to increase the magnetic
field, compression is negligible here because the initial
field as well as the infalling gas are radial (e.g., 
see the top right panel in Fig. 2).
The fast growth of the magnetic fields is exemplified
by the growth of the toroidal field as shown in the bottom row
of panels in Fig. 1.
We have verified that most of the inner torus is unstable to MRI, and
that our simulations have enough resolution to resolve, albeit
marginally, the fastest growing MRI mode (see also Section 3.2 
for the presentation of the results from run~B).  

The magnetic field effects drive the time evolution
of the torus including the mass accretion onto a BH. 
Another important effect of magnetic fields is that the torus produces 
a magnetized corona and an outflow (e.g., the middle panels in Fig. 1). 
The presence of the corona and outflow is essential to the
evolution of the inner flow at all times and the entire flow close to
the rotational axis during the latter phase of the evolution.  
The outflow very quickly becomes sufficiently strong to overcome
supersonically infalling gas (the radial Mach number in the polar
funnel near the inner radius is $\sim 5$) and makes its way outward,
reaching the outer boundary at $t=0.25$ s (see below and fig. 3). 
Due to limited computing
time, run~A was stopped at $t=0.28215$~s, which corresponds
to 6705 orbits of the flow near the inner boundary.  We
expect the accretion to continue much longer, roughly the collapse
timescale of the Helium envelope ($\sim 10$~s), as in MacFadyen \& 
Woosley (1999) .

Figure~2 shows the time evolution of several gross properties
run A: the mass accretion rate through
the inner boundary (top left panel), total magnetic energy (top right
panel), neutrino luminosity (bottom left panel) and radial Poynting and
kinetic flux along the polar axis at $r=190~R_S$ (bottom right panel).
Unless otherwise stated, all quantities in this paper are in cgs
units.  

Initially, during a precollapse phase, $\MDOT_a$ stays nearly
constant at the level of $\sim 5\times 10^{32}~{\rm g~s^{-1}}$. During
this phase the zero-$l$ gas inside the initial helium envelope is
accreted.  Around $t=0.13$~s, $\MDOT_a$ rises sharply as the gas from
the initial helium envelope reaches the inner boundary. However, this
gas has non-zero $l$ and a rotational supported torus and its corona
and outflow form (as illustrated in Fig.~1), 
causing a drop in $\MDOT_a$ after it reaches a
maximum of $2\times 10^{33}~{\rm g~s^{-1}}$ at $t=0.145$~s. 

The accretion rate reaches a minimum of $6\times10^{31}~{\rm g~s^{-1}}$ at
$t\approx0.182$~s and then fluctuates with a clear long-term increase.
This increase is caused by the contribution from gas with $l< 2 R_S
c$, which is directly accreted (without need to transport $l$) from
outside the main body of the torus (flow component C in Fig. 4 exemplifies 
a low $l$ inflow outside the torus). The total mass and angular
momentum accreted onto the BH during run~A (0.3 s) are
$0.1~\MSUN$ and $3\times10^{39}~{\rm g~cm^2~s^{-1}}$, respectively.

The top left panel of Fig.~2 shows the time evolution of the total magnetic 
energy (integrated over the entire computational domain). The late phase 
of the time evolution of the magnetic energy is characteristic of weakly 
magnetized rotating accretion flows. Most of the magnetic energy is due to 
the toroidal component of the field. We note a huge increase of
the toroidal magnetic field coinciding with the formation and
development of the torus. Both $B_\phi$ and $B_\theta$ are practically
zero while $B_r$ is constant during the precollapse phase of the evolution. 
But at $t=0.14$~s the total energy in $B_\phi$ equals that in $B_r$ 
and just 0.025~s later the $B_\phi$ energy is higher than the $B_r$ 
energy by a factor of 50. At the end of simulations the total kinetic energy 
from the radial, latitudinal and rotational motion are $4\times 10^{50}$,
$6.5\times 10^{49}$, and $2.3\times10^{51}$~erg, respectively.  These
gross properties indicate that the magnetic energy is large enough to
play an important role in the flow dynamics.

The bottom left panel of Fig.~2 shows
the time evolution of the neutrino luminosity, $L_\nu$. 
The neutrino luminosity was computed
under the assumption that all the gas in the model is
optically thin to neutrinos. 
The neutrino emission stays at a relatively constant level of
$3\times10^{52}~{\rm erg~s^{-1}}$ after the torus forms.
This indicates that the gross properties -- such as
the temperature, density, and total mass -- of
the densest and hottest parts of the flow (the torus) 
also stay constant during the late phase of the evolution.

The bottom right panel in Fig.~2 shows the area-integrated radial
fluxes of  magnetic and kinetic energy at $r=190~R_S$ inside
the polar outflow. 
Formally, the polar
outflow is defined as the region where $v_r>0$ and $\beta<1$.
The Poynting fluxes, stays nearly 
constant at the level of $\sim 1\times 10^{51}~{\rm egr~s^{-1}}$
between $t=0.185$~s
(the time when the jet reached the radius of 190~$R_S$)
and $t=0.22~s$
then fluctuates with a clear long-term decrease. Comparing the
two fluxes, we conclude that
the outflow is Poynting flux-dominated, with the Poynting
flux exceeding the kinetic energy flux by up to an order of magnitude.
We note that the radial fluxes continue to vary with time by a 
factor up to 10 even during the late phase of the evolution
and they are anti-correlated with the mass accretion.
(The radial fluxes show much less detail compared
to other properties shown in Fig.~2 because there were
computed less frequently during the course of the simulation.)

The decrease of the radial fluxes and their anti-correlation
with $\MDOT_A$ is quite a surprising result because
the jet is supposed to be powered by accretion! However, 
the situation in a collapsing star is complex because $\MDOT_a$ 
has two sources: (i) an accretion disk which does power a strong jet 
and (ii) the low $l$ gas that accretes directly onto a BH and
can prevent development of the torus jet.  This infalling gas
can produce an outflow/jet of its own but 
with properties different from those of the torus jet (see below).

Figure~3 shows maps of the flow pattern and  maps of $B_\phi$
on four different length scales at $t=0.285$~s (the left panels
show the inner most part of the flow whereas the right panels
show the flow on the largest scale).
The main outflow is magnetically driven from the torus.  
Soon after the torus forms, the magnetic field
very quickly deviates from its initial radial configuration due to MRI
and shear (see Fig. 2).  
This leads to fast growth of the toroidal magnetic field
as field lines wind up due to the differential rotation.  As a result
the toroidal field dominates over the poloidal field and the gradient
of the former drives an outflow.  

A comparison between the density and $B_\phi$ maps shows
that the polar regions of low density and high $B_\phi$ coincide with 
the region of an outflow. Proga  et al. (2003) 
noted that during the latter phase of the evolution not all of the
material in the outflow originated in the innermost part of the torus
-- a significant part of the outflow is ``peeled off'' the infalling
gas at large radii by the magnetic pressure.  However, as our later
analysis and simulations show the outflow from the infalling gas is 
caused not only by the peeling off effect but it is  also  
produced by the magnetic and thermal effects operating inside 
the infalling gas itself. Thus there are not two but
three types of outflows from a collapsing star (see Fig. 4).

The first two panels of the left hand side of Fig.~3 
illustrate that the inner torus and its corona and outflow cannot
always prevent the low-$l$ gas from reaching the BH.  Even the
magnetic field cannot do it if the density of the incoming gas is too
high (compare the upper and lower halfs of  the panels in Figs~3 and 5).

We finish the presentation of the results from run A with an analysis
of the properties of the accretion torus, i.e., the main engine
in the model. Figure~4 shows the radial profiles
of several quantities in run~A, angle-averaged over a small wedge near
the equator (between $\theta=86^\circ$ and $94^\circ$), and time-averaged
over 50 data files covering a period at the end of the simulations
(from 0.2629~s  through 0.2818~s).
We indicate the location of the last
stable circular orbit by the vertical dotted line in each panel.

We measure the Reynolds stress,
$\alpha_{gas}\equiv<\rho v_r \delta v_\phi>/P$,
and the Maxwell stress normalized to the magnetic pressure,
$\alpha_{mag}\equiv <2B_rB_\phi/B^2>$.
Note that Fig.~4 shows only the magnitude,
not the sign, of the normalized stresses.
We find that except for $r\simless 2.5 R_S$ and
$10 R_S\simless r \simless 12 R_S$,
the Maxwell stress dominates over the Reynolds
stress in the inner flow.
The last panel in Fig.~4 shows that the toroidal component
of the magnetic field is dominant for $r< 50~R_S$.

Proga et al.'s simulation explored a relatively conservative case 
as they allowed for neutrino emission but did not allow for 
the emitted neutrinos
to interact with the gas or annihilate.  The only sources of
nonadiabatic heating in their simulation are the artificial viscosity
and resistivity.

\subsection{High resolution run}

To explore the nature of the multicomponent flow resulting from
a collapse of a rotating star we rerun model~A using higher numerical 
resolution [see Section 3.1 and Proga (2005 in preparation)]. 
Higher resolution simulations are especially required to study
the evolution and effects of MRI. Run B  is qualitatively consistent with 
the results from the fiducial model. In particular, all the flow
components found in run~A can be also found in run~B. Additionally,
the properties of the flow components found in run~A are consistent
with those in run~A. For example, MRI drives the evolution of
the equatorial torus and the simulations resolve 
the fastest growing MRI mode.

Figure~5 shows the inner most part of the flow in run~B at t=0.582~s.
The seven main flow components are marked with upper case letters
(see the figure caption).

In the context of GRBs, the most important component of the 
flow is an outflow. The biggest difference between
the MHD collapsar model and the HD collapsar model considered 
by MacFadyen Woosley (1999) is that in the MHD limit, the outflow that develops
soon after a torus forms is so strong that it breaks through a star even 
when very little of a low $l$ gas is accreted whereas in the HD limit, a torus
wind is relatively weak and the polar funnel must be evacuated if the wind 
were to leave the star.

\subsection{An outflow from a very $l$ infall}

Generally, runs A and B show that large-scale magnetic fields can produce 
two types of outflows: (1) 
a jet from a rotationally supported accretion disk or torus
and (2) an outflow from extremely low angular momentum gas 
that almost radially accretes onto a BH (e.g., PB03, Proga et al. 2003).
Both analytic and numerical studies support the view that the former is robust.
(e.g., Blandford \& Payne 1982; Blandford 1990; De Villier et al. 2003). 
In fact, most MHD simulations of an accretion disk or torus show outflows
(e.g., Uchida \& Shibata 1985; Stone \& Norman 1994; Hawley \& Balbus 2002; 
De Villier et al. 2004; Mizuno et al. 2004; Kato et al. 2004; 
McKinney \& Gammie 2004). However, the latter which is 
a simple, self-consistent solution for the MHD jet problem, 
has not been studied much. 
To articulate the basic physics that occurs in jet production, Proga (2005)  
performed simulations of a flow with angular momentum so low that, if not 
for the effects of MHD, the flow would accrete directly onto a BH without 
forming a disk. These simulations used simplified physics (i.e., no neutrino 
cooling and an adiabatic equation of state) similar to that explored by PB03.

In Proga (2005), we found that even with a very weak initial magnetic field, 
the flow settles into a configuration with four components: 
(i)   an equatorial inflow,
(ii)  a bipolar outflow, 
(iii) a polar funnel outflow, and
(iv)  a polar funnel inflow.
The second flow component of the MHD flow represents a simple yet robust 
example of a well-organized inflow/outflow solution to the problem
of MHD jet formation (see Fig.~6) and is the same in nature as components 
C and B in run~B as well as in run~A (e.g., Fig.~5). 
The outflow from the low $l$ infall 
is heavy, highly magnetized, and  driven by magnetic and centrifugal forces. 
A significant fraction of the total energy in the jet is carried out 
by a large scale magnetic field. The properties of this outflow help to 
understand that the time evolution of the outflow in run~A.
As we mentioned in section~3.1, the low $l$ inflow can
block or prevent development of the torus jet, the latter being light.  
The low $l$ inflow produces the outflow that is
heavy  and although a significant fraction of its total energy 
is carried out by a large scale magnetic field it is not Poynting
flux-dominated, contrary to the torus jet.

A comparison between various simulations including those presented
in section 3.1 and 3.2, where specific angular momentum was higher than 
that assumed in Proga (2005), indicates that the flow components B and C 
develop for a wide range of the properties of the flow near the equator
and near the poles. 

\section{Conclusions}

Fully 3 dimensional GR MHD simulations are required to capture many of 
the effects and instabilities of a magnetized fluid in a rotating star 
collapsing onto a BH . Neutrino physics, a sophisticated equation 
of state and self-gravity should also be included. However, such full 
treatment of the MHD collapsar model is beyond reach of the current numerical 
codes at least for now.

Here we present results from time-dependent two-dimensional 
MHD simulations of the collapsar model using pseudo-Newtonian potential.  
The simulations show that: 
1) soon after the rotationally supported torus forms, the magnetic field very 
quickly starts deviating from purely radial due to MRI and shear. 
This leads to fast growth of the toroidal magnetic field as field lines 
wind up due to the torus rotation; 
2) The toroidal field dominates over the poloidal field and the gradient 
of the former drives a torus outflow against supersonically accreting 
gas through the polar funnel; 
3) The torus outflow is Poynting flux-dominated; 
4) The torus outflow reaches the outer boundary of the computational 
domain ($5\times10^8$~cm) with an expansion velocity of 0.2 c; 
5) The torus outflow is in a form of a relatively narrow jet 
(when the jet breaks through the outer boundary its half opening angle is 
$5^\circ$); 
6) Most of the energy released during the accretion is in neutrinos, 
$L_\nu=2\times 10^{52}~{\rm erg~s^{-1}}$. Neutrino driving will increase 
the outflow energy (e.g., Fryer \& M\'{e}sz\'{a}ros 2003 and references 
therein), but could also increase the mass loading of the outflow 
if the energy is deposited in the torus. 
A comparison of the MHD simulations with their HD counterparts show that 
a strong outflow breaks through a magnetized star much sooner than through 
a non-magnetized star. 

The above conclusions were reached by Proga et al. (2003) and here we 
confirmed them using a higher resolution simulation. However, 
we emphasize the fact that the flow settles into a complex convolution 
of several distinct, time-dependent flow components and the above mentioned 
torus and its outflow are just two of them. Other flow components include 
a torus corona and low $l$ flows. A rotationally 
supported torus and its corona and outflows were extensively studied in 
the past and remain a focus of many studies. We stress that in the context
of the collapsar model, where very low $l$ and high $l$ fluids are present,
the situation is more complex. Therefore, future work, even 
without neutrino physics or effects of general relativity, is important
to explore connection and interaction of all the flow components, and 
their observational implications.

\begin{figure*}
\begin{picture}(180,590)
\put(-220,323){\includegraphics{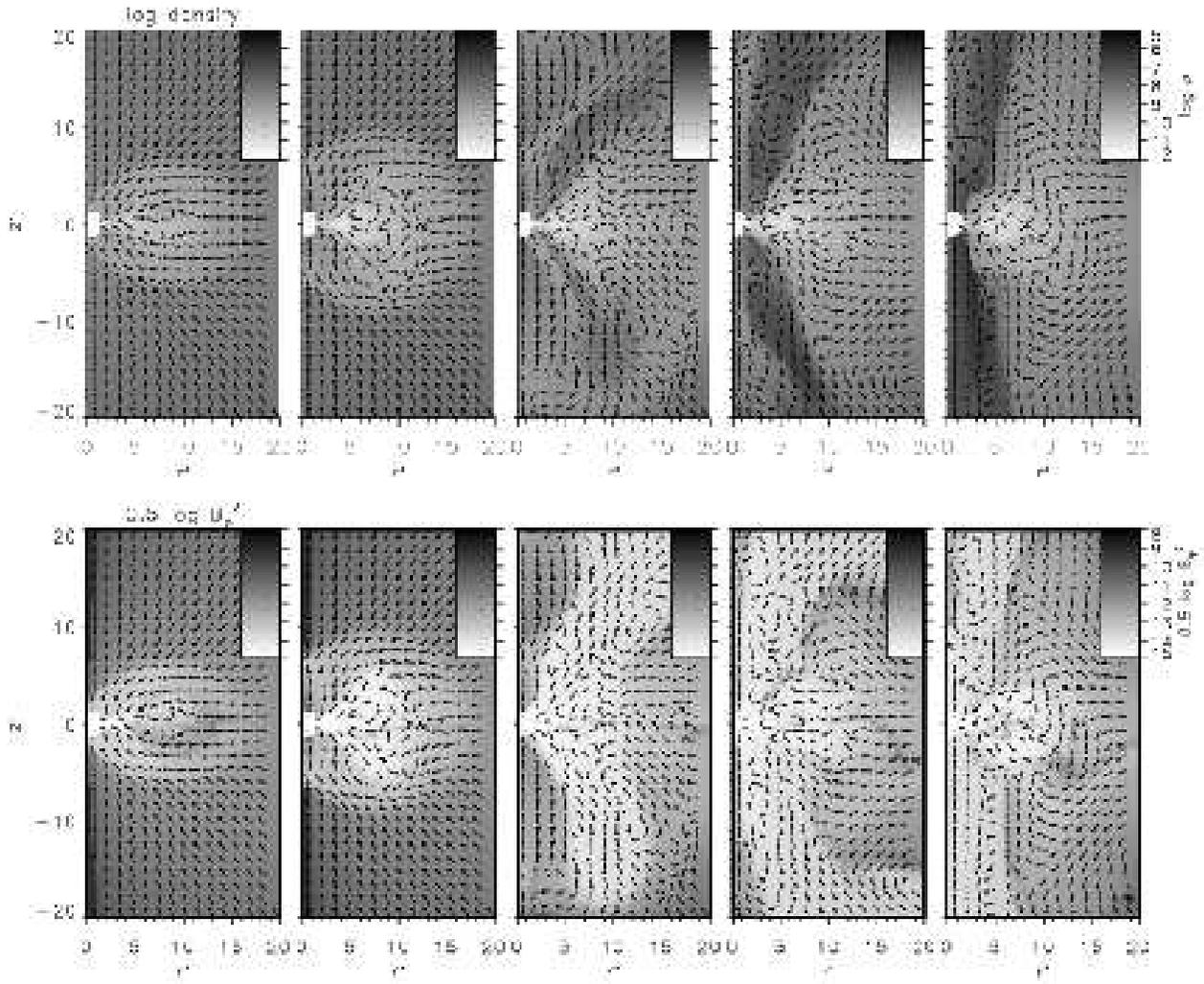}}










\end{picture}
\caption{
Sequence of  logarithmic density (top) and  toroidal magnetic field maps
(bottom) overplotted with the direction
of the poloidal velocity from run A at times 0.153, 0.161, 0.165, and 
0.173 s. The sequence illustrates the early phase of the formation
of a rotational supported accretion torus and 
and of a magnetically driven outflow.
The length scale is in units of the BH radius (i.e., $r'=r/R_S$
and $z'=z/R_S$).
}
\end{figure*}

\begin{figure*}
\begin{picture}(280,500)
\put(140,500){\includegraphics{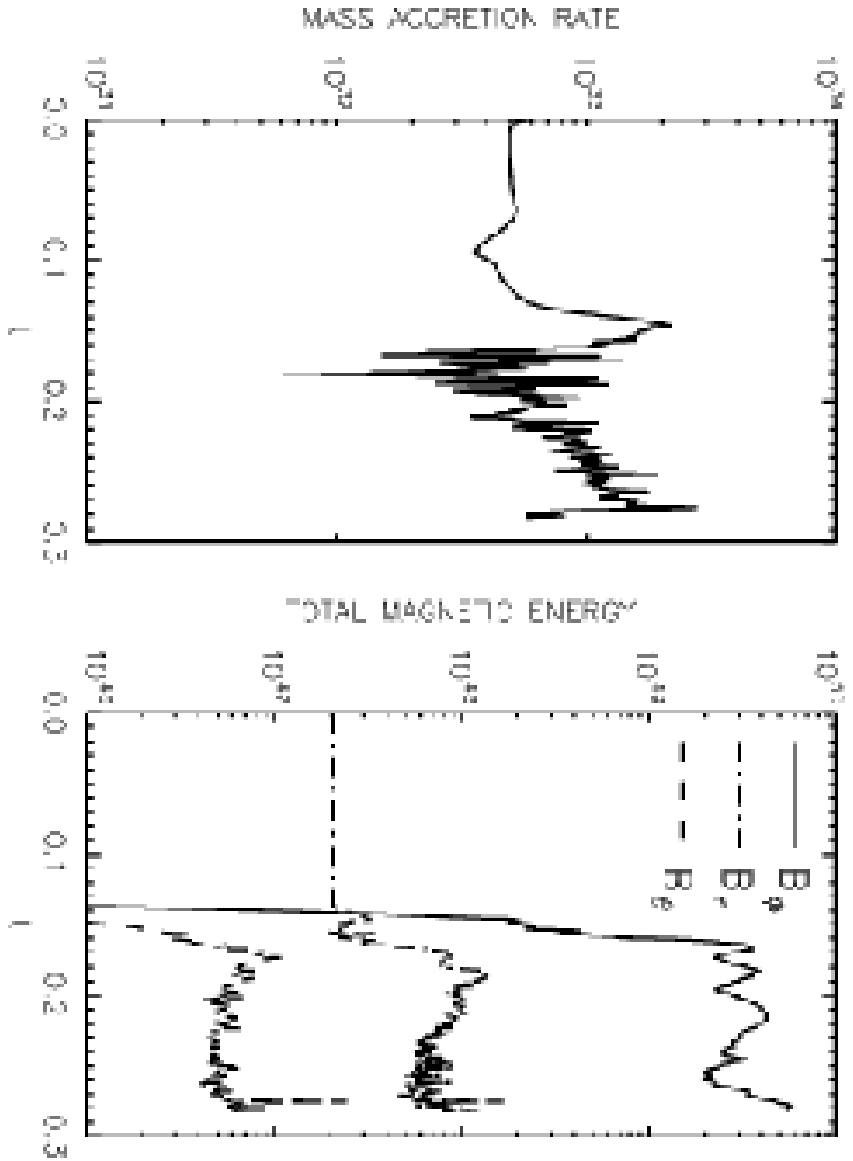}}
\put(148,250){\includegraphics{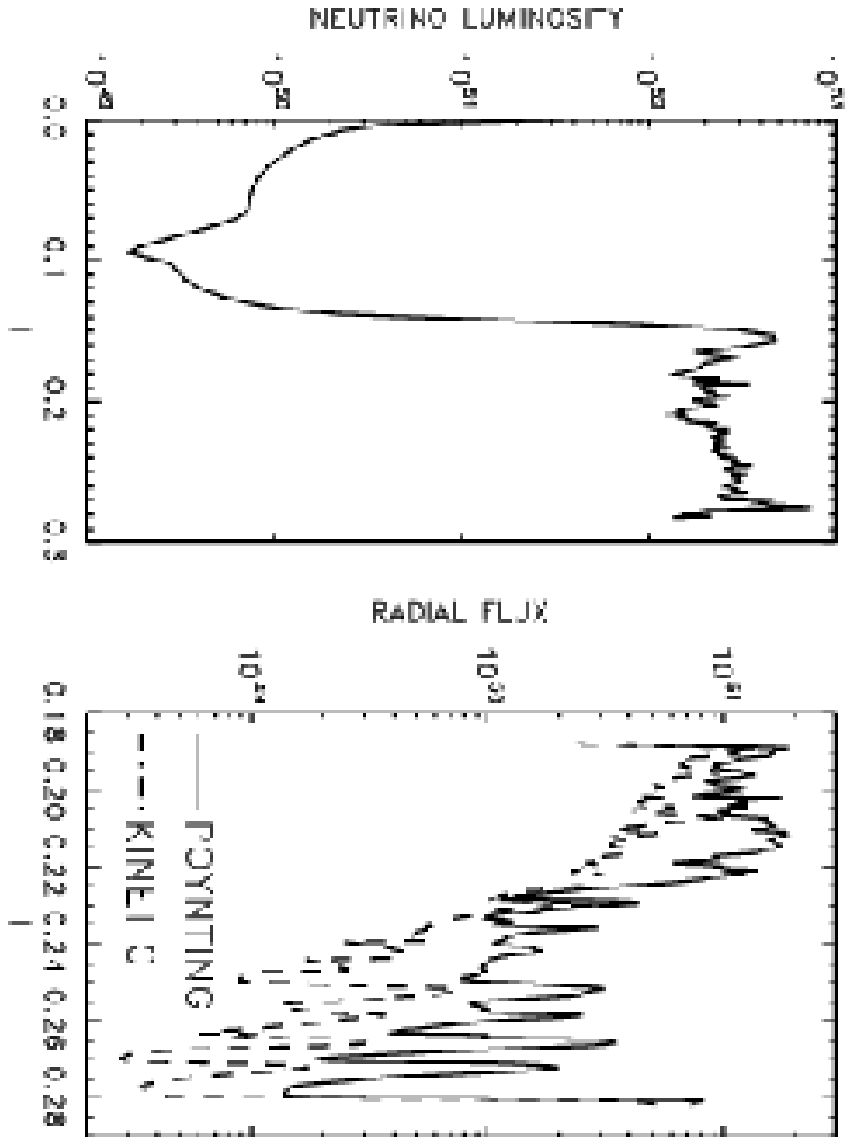}}
\end{picture}
\caption{
The time evolution of the mass accretion rate (top left panel),
total magnetic energy due to each of the three field components
(top right panel), neutrino luminosity 
(bottom left panel) 
and area-integrated radial Poynting and kinetic flux in the polar outflow at
$r=190~R_S$ (bottom right panel) for run A. 
Formally, we define the polar outflow as the region where $v_r>0$ and 
$\beta<1$. Note the difference in the time range in the panel with the
radial fluxes.
}
\end{figure*}

\begin{figure*}
\begin{picture}(180,590)
\put(-220,523){\includegraphics{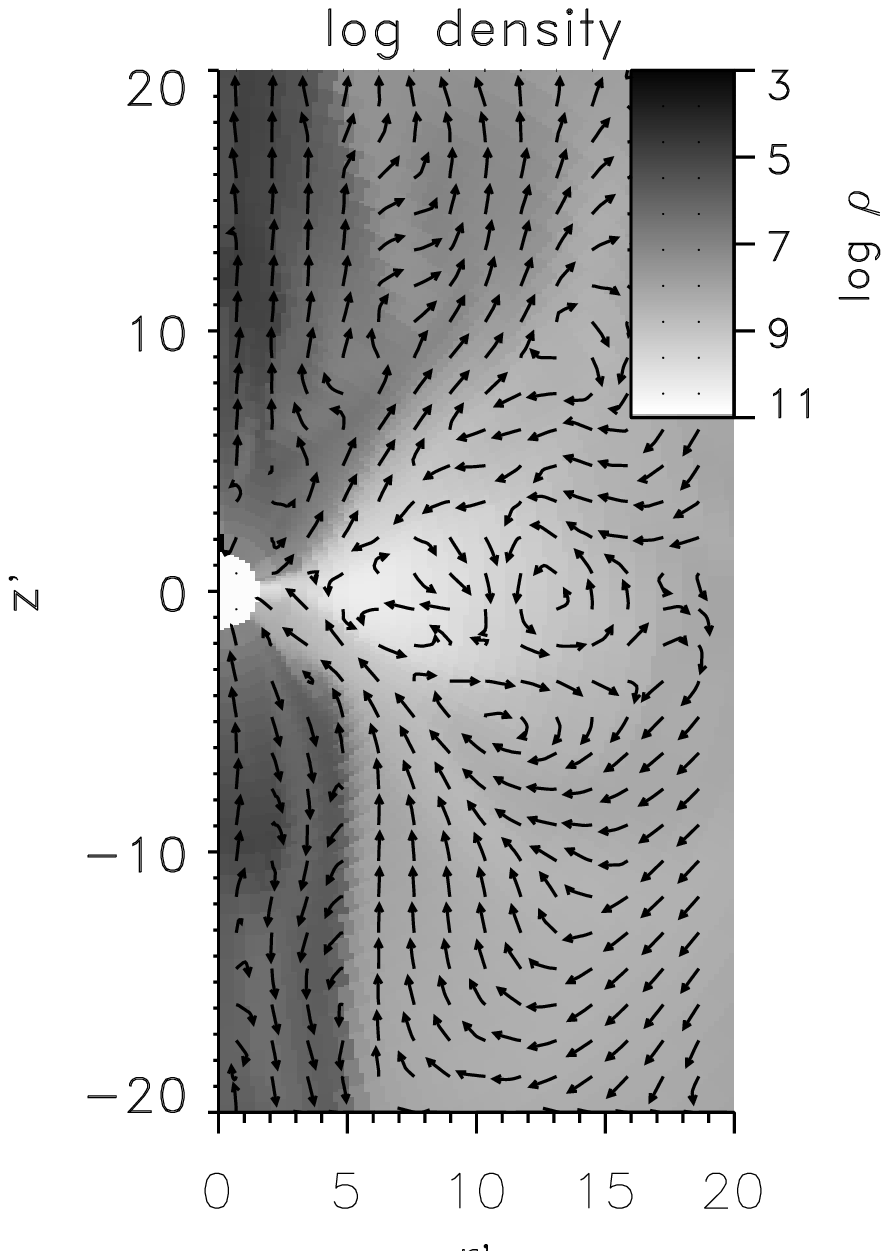}}

\put(-90,523){\includegraphics{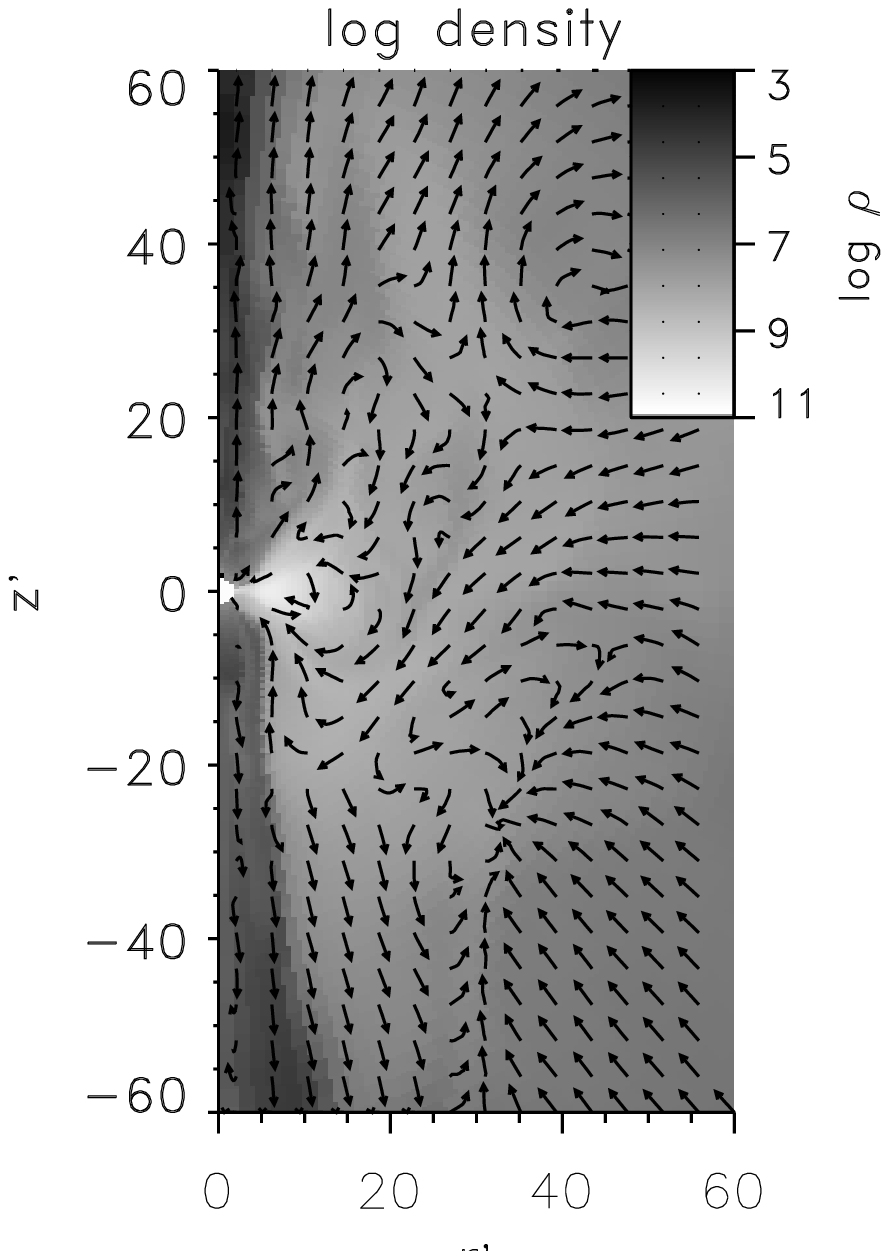}}

\put( 40,523){\includegraphics{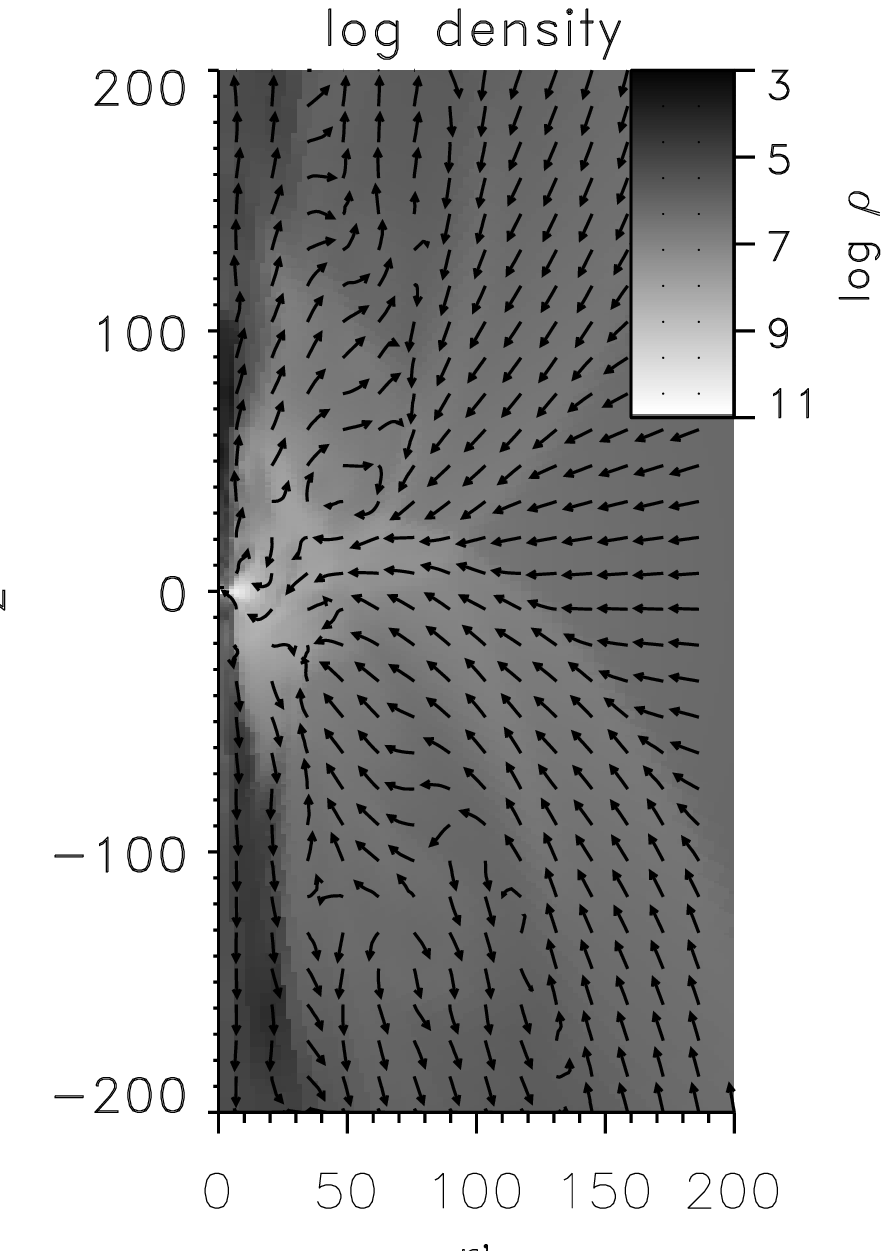}}

\put(180,523){\includegraphics{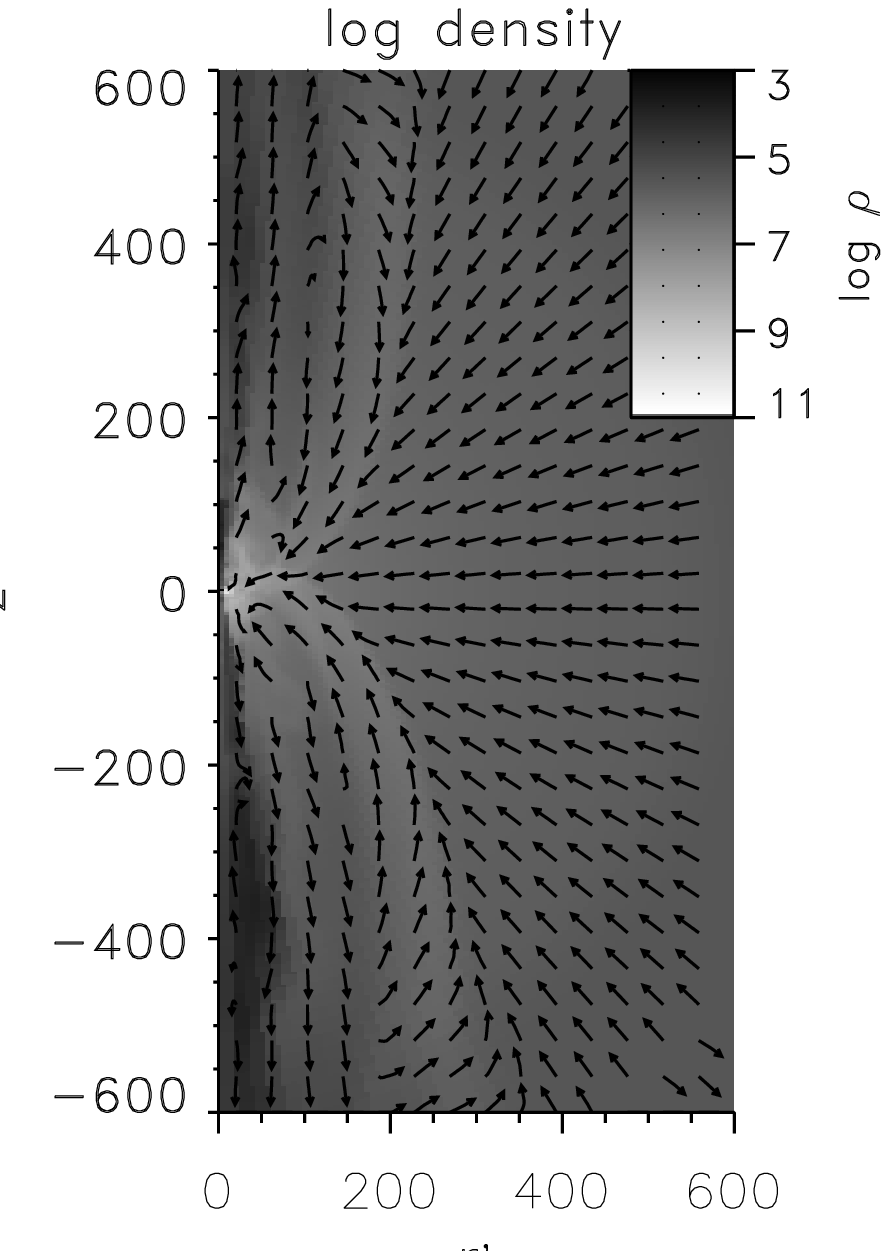}}

\put(-220,290){\includegraphics{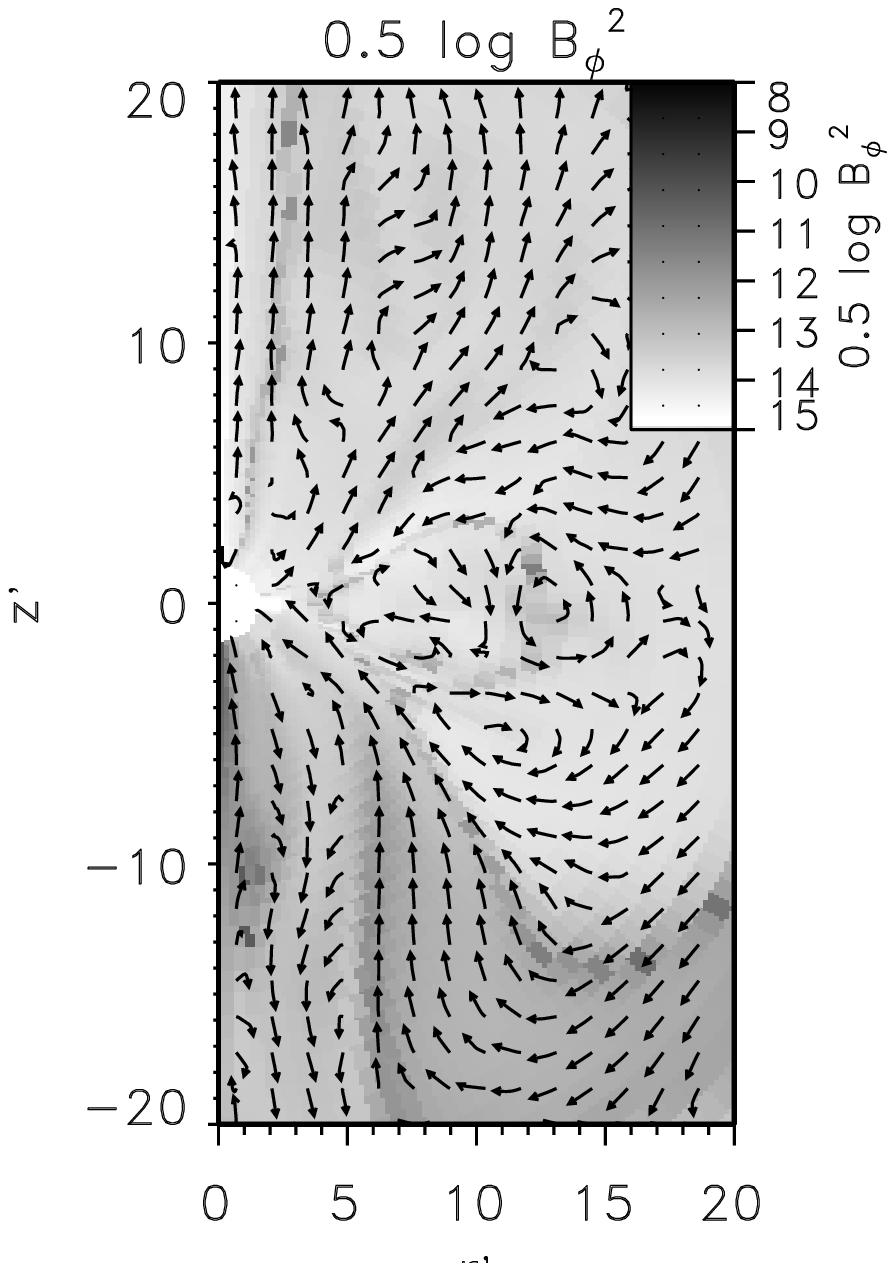}}

\put(-90,290){\includegraphics{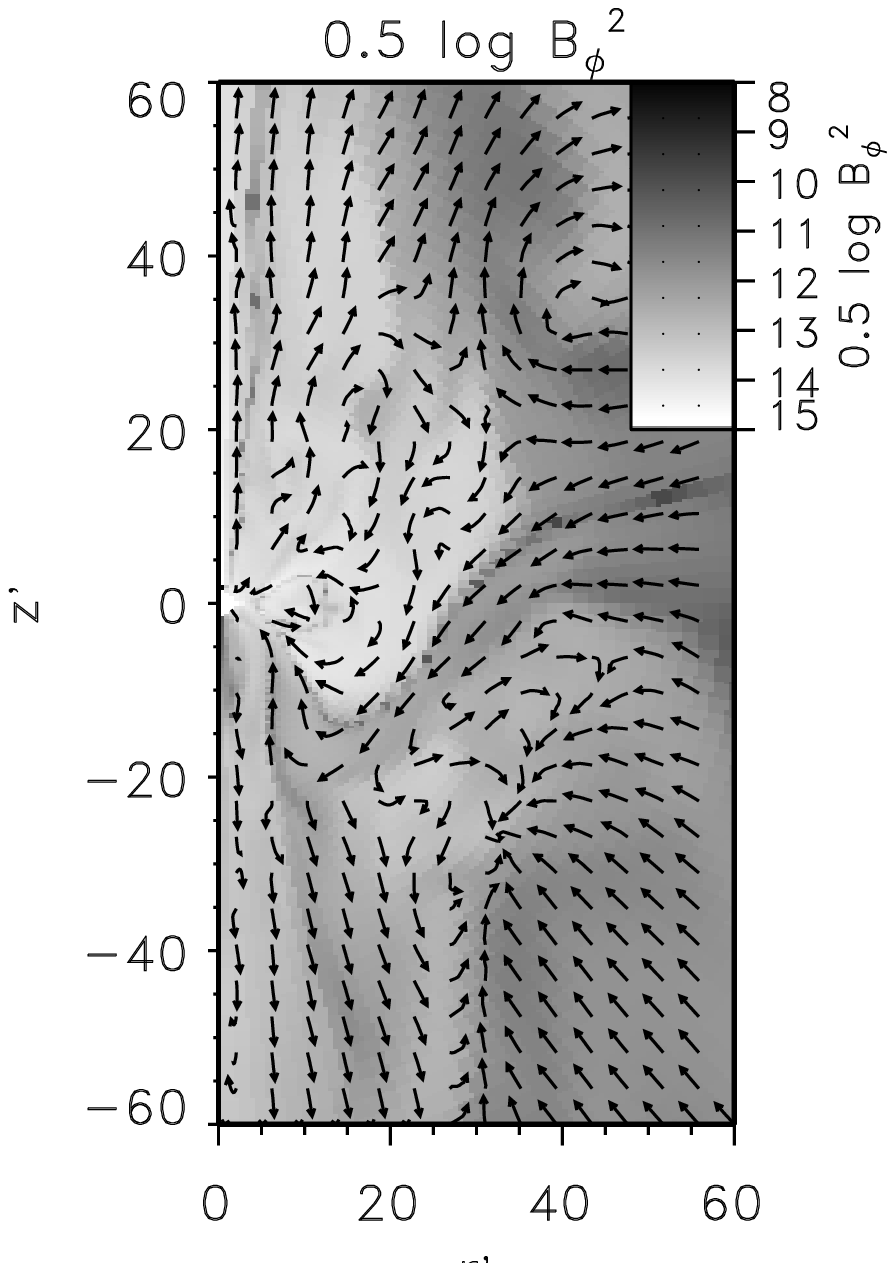}}

\put( 40,290){\includegraphics{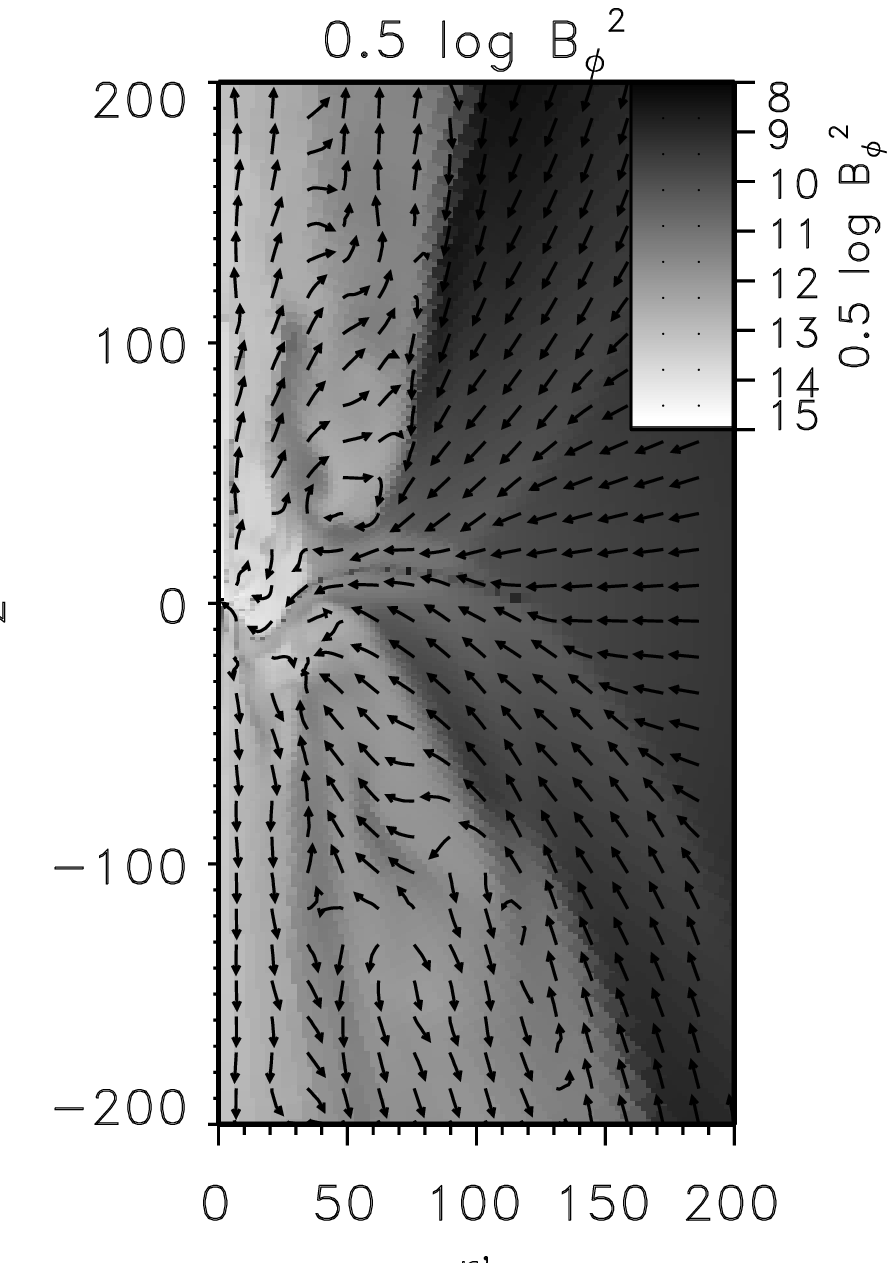}}

\put( 180,290){\includegraphics{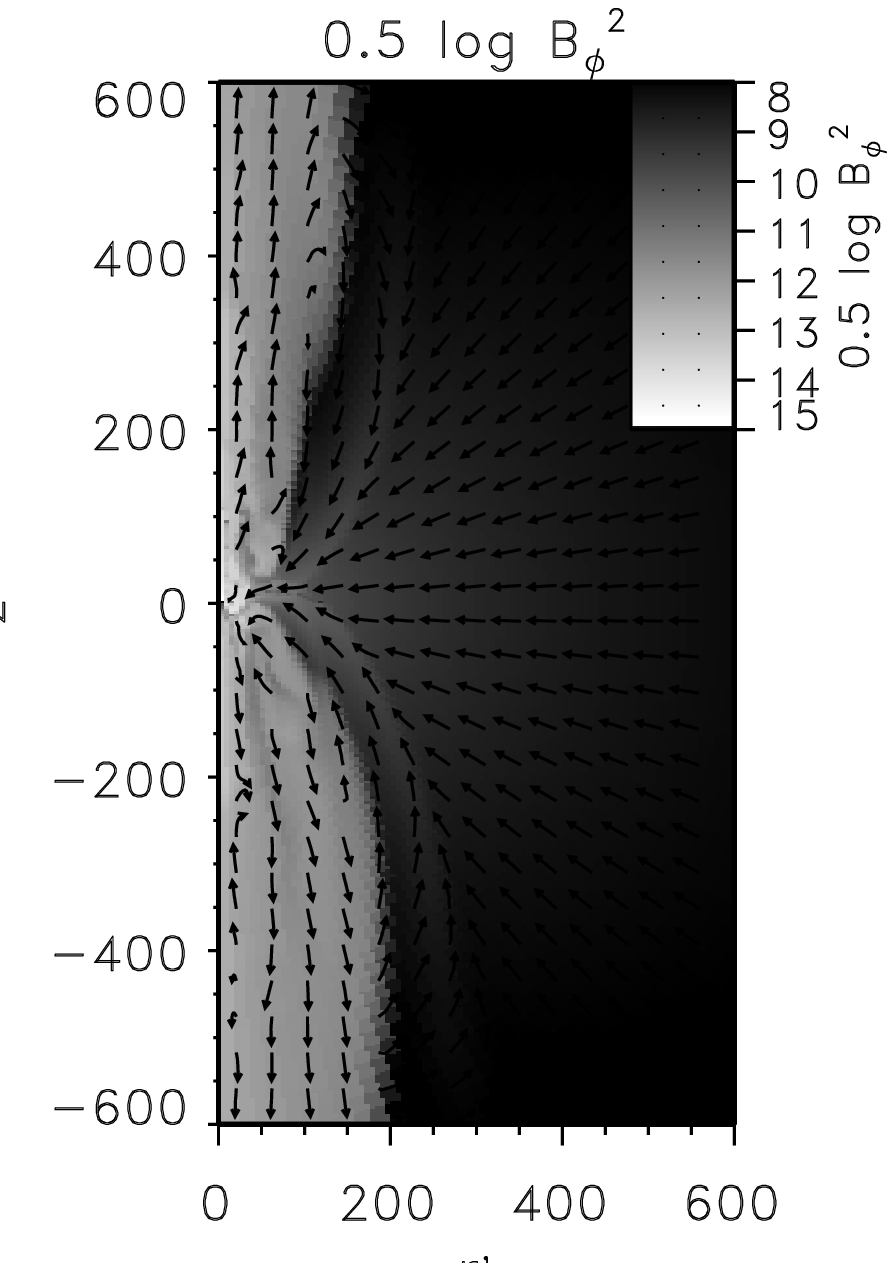}}

\end{picture}
\caption{
Maps of logarithmic density (top) and  toroidal magnetic field maps
(bottom) overplotted with the direction
of the poloidal velocity from run A at the end of simulations, i.e.
time 0.2815 s. The length range increases from to the right.
Note that the accretion torus is relatively small (it spans from
1 to about 20 $R_S$). Nevertheless, this tiny torus generates
an outflow and mass and energy that can the dynamics and structure
of the collapsing star over a large range of radii along the rotational
axis.
}
\end{figure*}

\begin{figure*}
\begin{picture}(180,590)
\put(280,423){\includegraphics{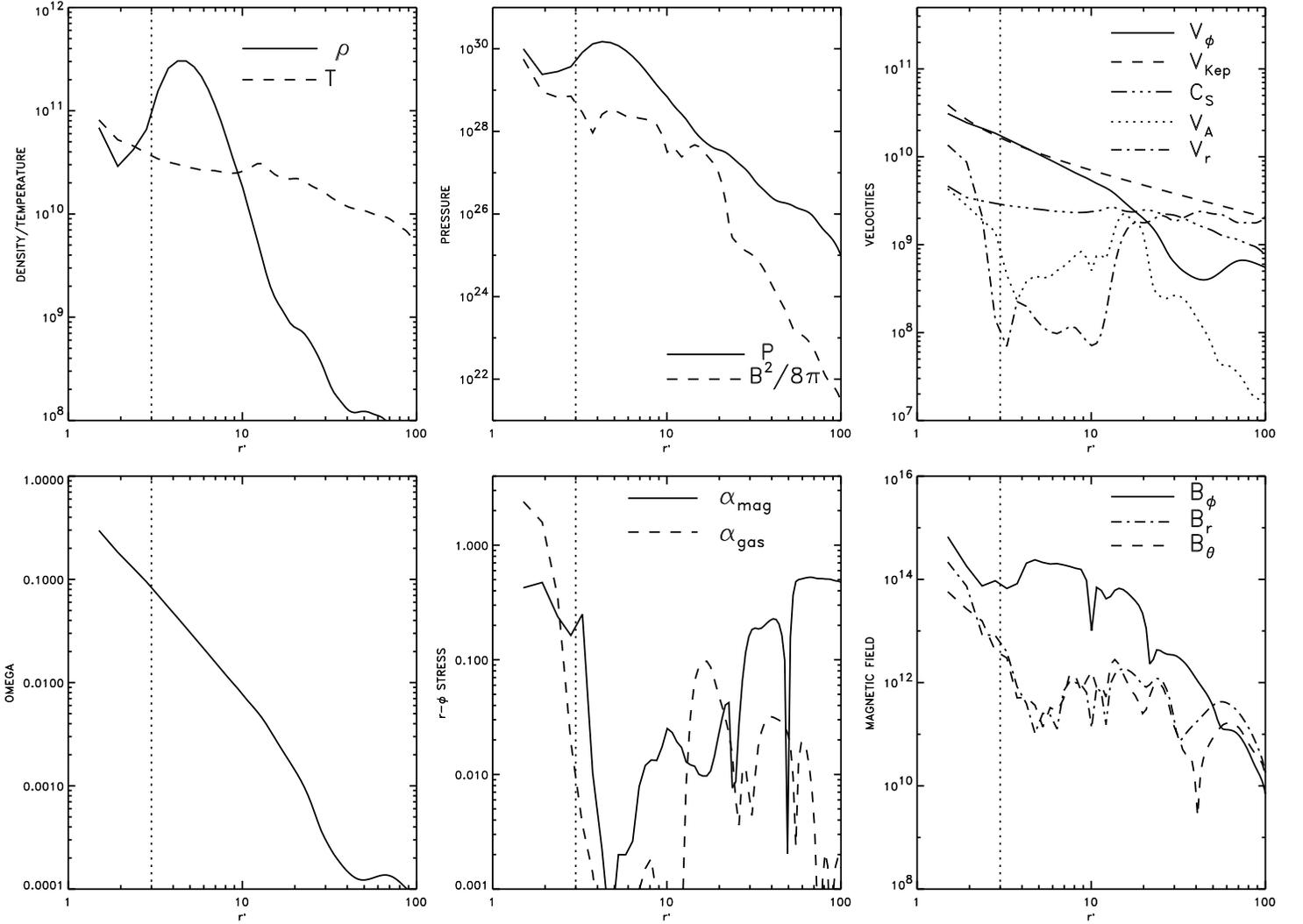}}
\end{picture}
\caption{Radial profiles of various quantities from our run, 
time-averaged 
from $0.2629$  through $0.2818$~s. 
To construct each plot, we averaged the profiles
over angle between $\theta=86^\circ$ and $94^\circ$.
The top left panel plots the density (solid line) and temperature (dashed line).
The top middle panel plots the gas pressure (solid line) and magnetic
pressure. The top right panel plots the rotational, radial,
Keplerian, and Alfv${\acute{\rm e}}$n velocities  
(solid, dashed, dot-dashed, and dotted line, respectively), as well as
the sound speed (triple-dot dashed line).
The bottom left panel plots the angular velocity in units of $2c/R_s$.
The bottom middle panel plots the Maxwell stress,  
$\alpha_{mag}$,
and the Reynolds stress, $\alpha_{gas}$
(solid and dashed line, respectively). 
We calculate the Reynolds stress using eq. (15) in PB03 
and  show only its amplitude.
The bottom right panel plots
the radial, latitudinal and toroidal components of the magnetic field
(dot-dashed, dashed, and solid line, respectively).
The length scale is in units of the BH radius (i.e., $r'=r/R_S$).}
\end{figure*}

\begin{figure*}
\begin{picture}(180,490)
\put(-220,300){\includegraphics{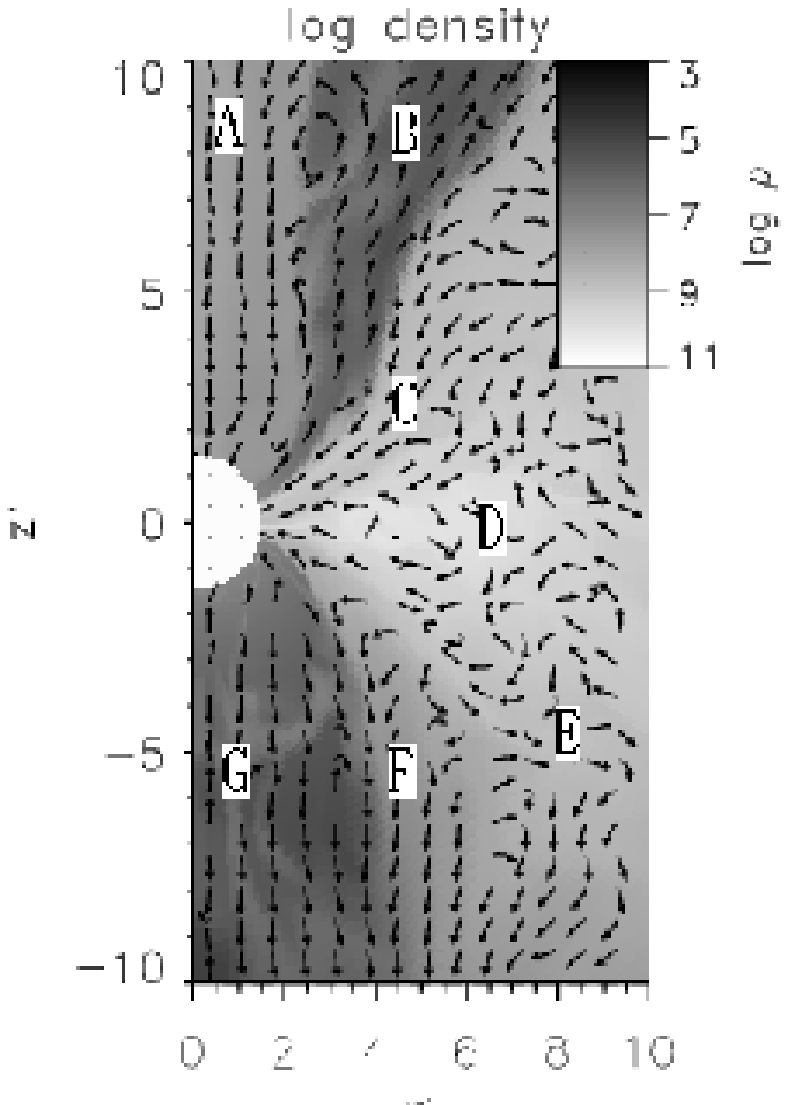}}
\put(-20,300){\includegraphics{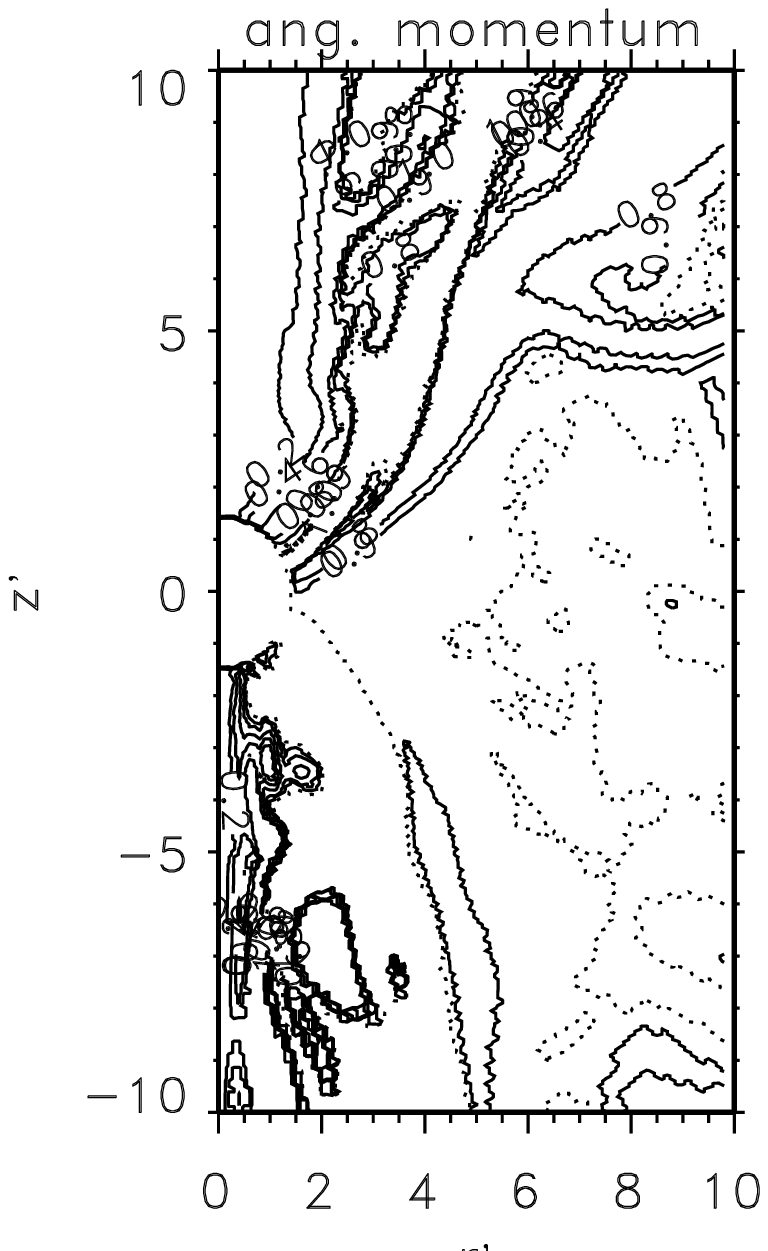}}
\end{picture}
\caption{
A map  of logarithmic density 
overplotted with the direction
of the poloidal velocity (left panel) and
a contour map of specific angular momentum, $l$ (right panel)
from high resolution run B at $t=0.0582$~s.
The specific angular momentum is in units of 2$R_S$c. 
The minimum of $l$ (contour closest to the rotational z-axis) is 0.2, 
and the contour levels are equally spaced at intervals of $l = 0.2$.
The maximum of $l$ is 1.0 and its contour is plotted using dotted curves, whereas all the other contours are plotted using solid curves.
This figure shows an inner most part of the flow when a torus
just formed and started to develop an outflow. Note that the latter 
pushed aside the polar funnel accretion flow only below the equator.
This figure was chosen to illustrate the complexity of
the inner most part of the flow inside a collapsing magnetized star.
The upper case letters, in the right panel, mark seven major components 
of this complex flow: 
A -- a very low $l$ polar funnel accretion flow;
B -- a highly magnetized outflow generated from a low $l$ inflow C;
C -- an inflow with angular momentum so low
that the flow would accrete directly onto a BH; 
D -- a rotationally supported, MHD, turbulent accretion torus;
E -- a magnetized torus corona;
F -- a highly magnetized outflow generated from the torus; and
G -- a polar funnel outflow driven thermally (compare with A). 
See Proga (2005; in preparation)
for more details.
}
\end{figure*}

\begin{figure*}
\begin{picture}(180,590)
\put(40,300){\includegraphics{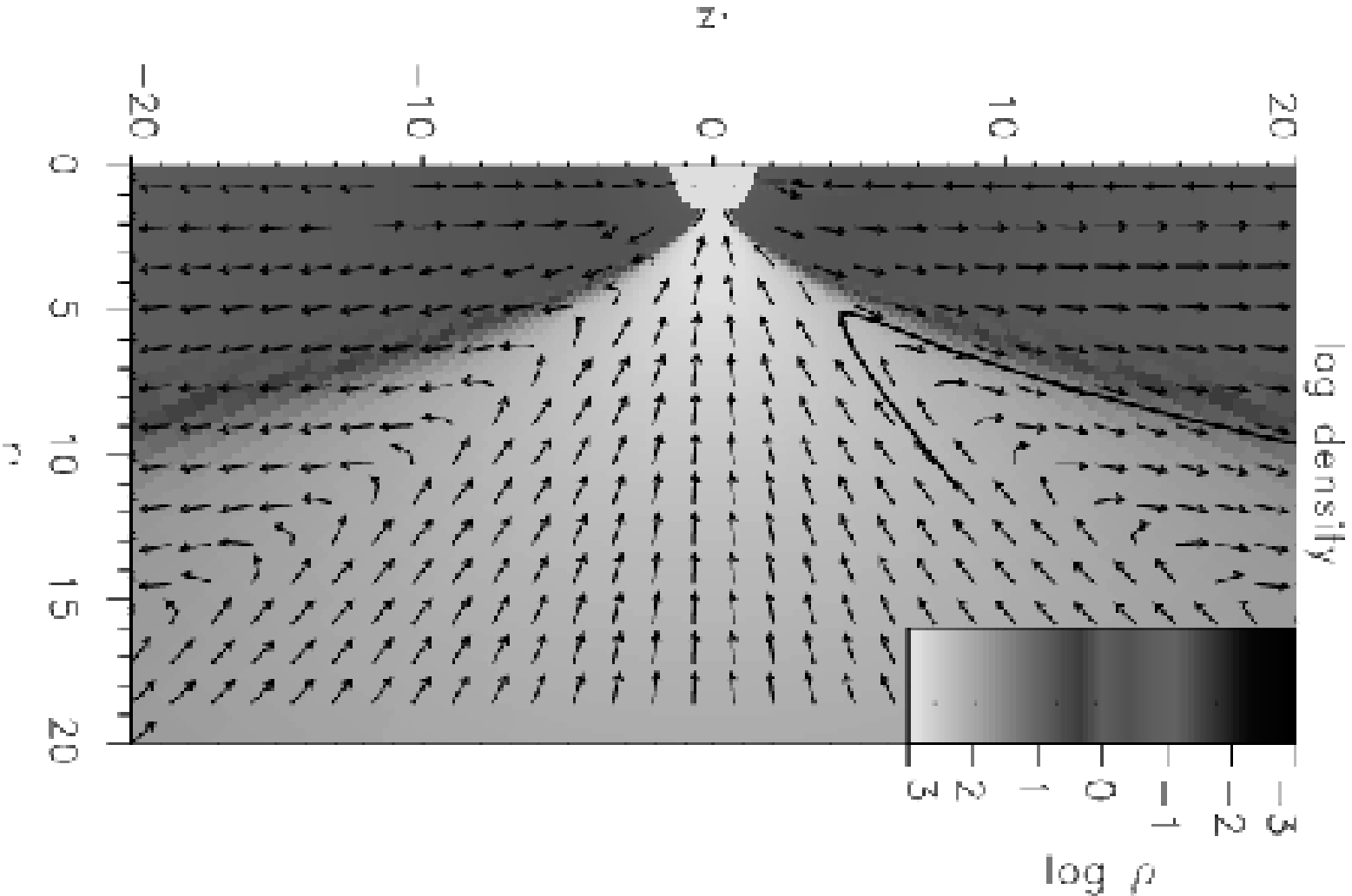}}
\put(300,300){\includegraphics{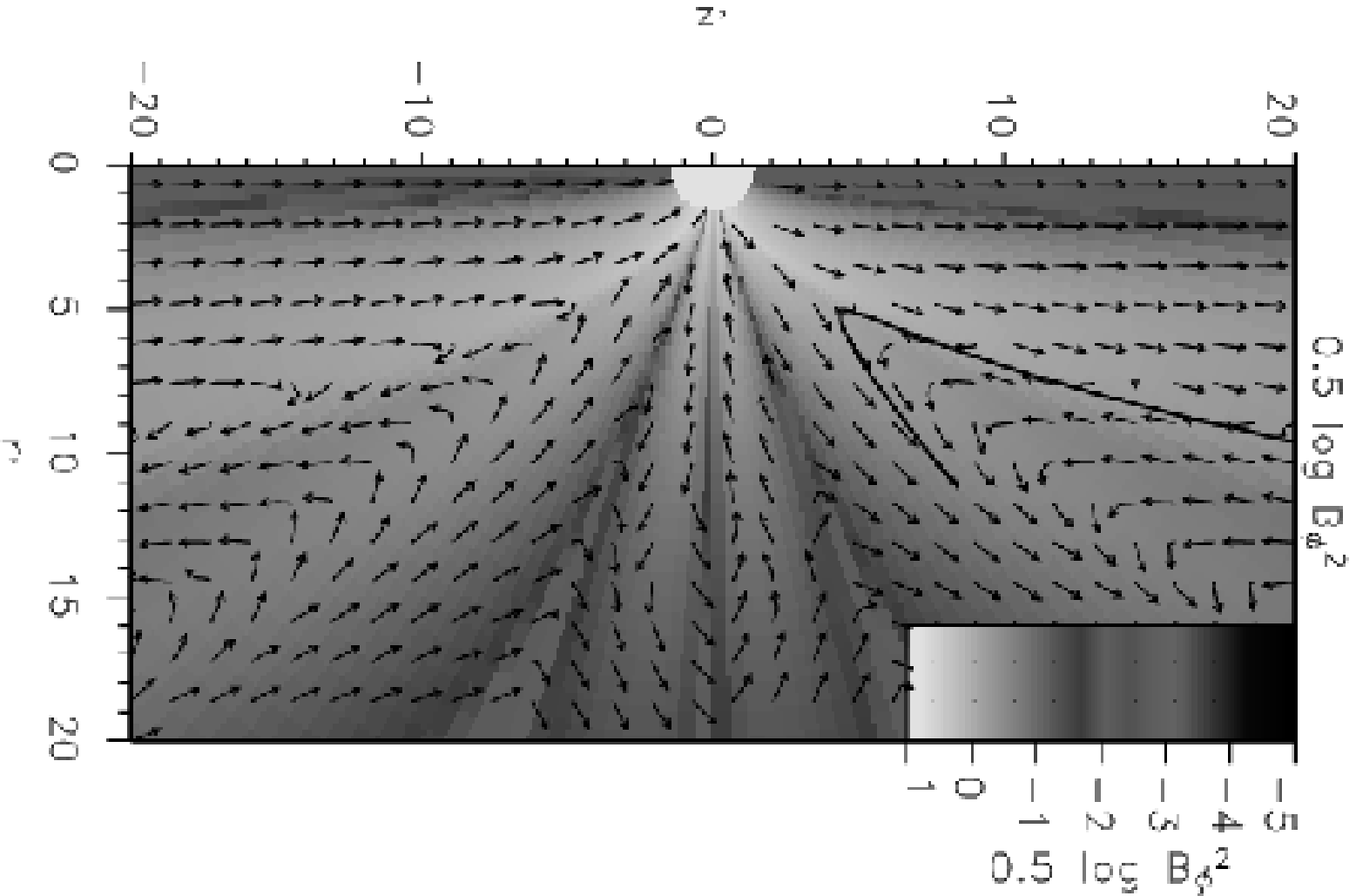}}
\end{picture}
\caption{
Maps of logarithmic density (left panel) and toroidal magnetic field 
(right panel) overplotted with an example of a streamline 
corresponding to an inflow/outflow for model~C. 
The maps are also overplotted with
the direction of the poloidal velocity and
the direction of the poloidal field 
(the left  and right panels, respectively). See Proga (2005) for more details.
}
\end{figure*}

\acknowledgments

We acknowledge support from NASA under ATP grant NNG05GB68G
and support provided by NASA through grant  HST-AR-10305.05-A
from the Space Telescope Science Institute, which is operated
by the Association of Universities for Research in Astronomy, Inc.,
under NASA contract NAS5-26555.

\end{document}